\documentclass[aps,prd,twocolumn,superscriptaddress,dvipdfmx]{revtex4}
\usepackage{amsmath,amssymb,amsfonts}
\usepackage{color}
\usepackage{graphicx}
\usepackage{enumerate} 
\usepackage{colordvi} 
\usepackage{bm}
\usepackage{hyperref}

\newcommand{\mpc}{\, {\rm Mpc}}

\newcommand{\hmpc}{\, h^{-1} \mpc}
\newcommand{\ihmpc}{\, h\, {\rm Mpc}^{-1}}

\newcommand{\vx}{\mathbf{x}}

\newcommand{\vk}{\mathbf{k}}

\newcommand{\himpc}{{\hbox {$~h^{-1}$}{\rm ~Mpc}}}
\newcommand{\hmpci}{{\hbox {$~h{\rm ~Mpc}^{-1}$}}}

\newcommand{\himsun}{{\hbox {$~h^{-1}$}{M_\odot}}}

\newcommand{\be}{\begin{equation}}
\newcommand{\ee}{\end{equation}}
\newcommand{\bey}{\begin{eqnarray}}
\newcommand{\eey}{\end{eqnarray}}

\newcommand{\nn}  {\nonumber}

\begin{document}
\title{
Galaxy power spectrum  in redshift space: \\ combining perturbation theory with the halo model
}
\author{Teppei Okumura}\email{teppei.okumura@ipmu.jp}
\affiliation{
Kavli Institute for the Physics and Mathematics of the Universe (WPI), The University of Tokyo Institutes for Advanced Study, The University of Tokyo, Kashiwa, Chiba 277-8583, Japan}

\author{Nick Hand}
\affiliation{Astronomy Department, University of California, CA 94720, USA}

\author{Uro{\v s} Seljak}
\affiliation{Astronomy Department, University of California, CA 94720, USA}
\affiliation{Lawrence Berkeley National Laboratory and Physics Department, University of California, CA 94720, USA}

\author{Zvonimir Vlah}
\affiliation{Stanford Institute for Theoretical Physics and Department of Physics, Stanford University, Stanford, CA 94306, USA }
\affiliation{Kavli Institute for Particle Astrophysics and Cosmology, SLAC and Stanford University, Menlo Park, CA 94025, USA}

\author{Vincent Desjacques}
\affiliation{D\'epartement de Physique Th\'eorique and Center for Astroparticle Physics (CAP), Universit\'e de Gen\`eve, 1211 Gen\`eve, Switzerland}

\date{\today}

\begin{abstract}
Theoretical modeling of the redshift-space power spectrum of galaxies is crucially important to correctly extract cosmological information from galaxy redshift surveys.
The task is complicated by the nonlinear biasing and redshift space distortion (RSD) effects, which change with halo mass, and by the 
wide distribution of halo masses and their occupations by galaxies. One of the main modeling challenges is
the existence of satellite galaxies that have both radial distribution inside the halos and large virial velocities inside halos, a phenomenon known as the Finger-of-God (FoG) effect. 
We present a model for the redshift-space power spectrum of galaxies in which we decompose a given galaxy sample into central and satellite galaxies and relate different contributions to the power spectrum to 1-halo and 2-halo terms in a halo model. 
Our primary goal is to ensure that any parameters that we introduce have physically meaningful values, and are not just 
fitting parameters. 
For the lowest order 2-halo terms we use the previously developed RSD modeling of halos in the context of distribution function and perturbation theory approach. This term needs to be multiplied by the effect of radial distances and velocities of satellites inside the halo. 
To this one needs to add the 1-halo terms, which are non-perturbative. 
We show that the real space 1-halo terms can be modeled as almost constant, 
with the finite extent of the satellites inside the halo inducing a small 
$k^2R^2$ term over the range of scales of interest,  where $R$ is related 
to the size of the halo given by its halo mass. 
We adopt a similar model for FoG in redshift space, ensuring that FoG velocity dispersion is related to the halo mass. For
FoG $k^2$ type expansions do not work over the range of scales of interest 
and FoG resummation must be used instead. We test several
simple damping functions to model the velocity dispersion FoG effect. 
Applying the formalism to mock galaxies modeled after the ``CMASS'' sample of the BOSS survey, we find that our predictions for the redshift-space power spectra are accurate up to $k\simeq 0.4 \hmpci$ within 1\% if the halo power spectrum is measured using $N$-body simulations and within 3\% if it is modeled using perturbation theory.
\end{abstract}
\pacs{98.80.-k}
\keywords{cosmology, large-scale structure} 
\maketitle

\flushbottom
\section{Introduction}

Redshift surveys enable us to probe the three-dimensional mass density field, while weak lensing surveys and cosmic microwave background experiments measure the density field projected along the observer's line of sight. 
The measured distances to galaxies are measured through redshift and are distorted due to radial components of peculiar velocities. These changes are known as redshift-space distortions (RSD) \cite{Kaiser:1987, Hamilton:1998} and contain additional cosmological information. Analyzing the power spectrum or correlation function in redshift space provides a useful test of dark energy and general relativity, e.g., \cite{Guzzo:2008,Jain:2008,Song:2009,McDonald:2009,White:2009} (see, e.g., \cite{Beutler:2014,Reid:2014} for the recent observational studies).
However, galaxy clustering is known to suffer from various kinds of nonlinear effects, and we need to model them in order to extract all possible information from redshift surveys, e.g., \cite{Fisher:1995, Heavens:1998, Scoccimarro:1999, Bharadwaj:2001, Scoccimarro:2004,Tinker:2006, Okumura:2011, Jennings:2011, Kwan:2012}. 
Nonlinearity in the power spectrum can be modeled using perturbation theory (PT) \cite{Bernardeau:2002} and numerical simulations \cite{Smith:2003,Heitmann:2010}.
The nonlinearity of RSD was first modeled for dark matter \cite{Matsubara:2008, Taruya:2009, Taruya:2010, Valageas:2011, Seljak:2011, Okumura:2012, Vlah:2012, Zhang:2013,Senatore:2014}, and the formalisms have been extended to dark matter halos \cite{Matsubara:2008a, Reid:2011, Sato:2011, Nishimichi:2011, Okumura:2012b, Carlson:2013, Vlah:2013, Blazek:2014, Song:2014a}. 
Although detailed studies are required to fully understand halo bias \cite{Chan:2012, Baldauf:2012, Biagetti:2014, Saito:2014}, the theoretical models for the redshift-space power spectrum of halos were shown to work well up to reasonably small scales. 

However, what one observes in real observations is the redshift-space power spectrum not of halos but of galaxies. 
Although all the galaxies are considered to be formed inside dark matter halos, modeling the galaxy power spectrum is much more complicated than the halo spectrum
because of the large virial motions of satellite galaxies, which is known as the Finger-of-God (FoG) effect \cite{Jackson:1972}. 
The FoG effect is a fully nonlinear process, caused by virialized motions of satellite galaxies inside the halos, 
and depends strongly on both the mass of the host halo and the satellite fraction \cite{White:2001, Seljak:2001}. 
It is not possible to model the effect in  PT schemes, but the effect can be related to the underlying halo mass and expected
satellite occupation.
Usually, PT resummed damping functions such as Gaussian or Lorentzian 
have been considered \cite{Peacock:1994, Park:1994, Scoccimarro:2004, Percival:2009, Taruya:2010} and multiplied by the halo spectrum to model the galaxy power spectrum in redshift space.
They contain at least one free parameter, the velocity dispersion of the halos in which satellites are. 
Assuming we know the form of the damping function, we must still consider the different halo masses that contribute to FoG. 
This is usually done in the context of the so-called halo model \cite{Seljak:2000, Peacock:2000, Ma:2000, Scoccimarro:2001, Cooray:2002}. The halo model has been adopted to the galaxy clustering in redshift space \cite{White:2001, Seljak:2001, Kang:2002, Yang:2004, Tinker:2007, Hikage:2012,Hikage:2014,Reid:2014}.
The halo power spectrum and correlation function were directly measured from $N$-body simulations in \cite{Hikage:2014,Guo:2015} and \cite{Reid:2014}, respectively, to fully take into account the nonlinearities of halo clustering in the halo model.
However, for analytical approaches linear PT has been used to describe RSD of halos, i.e., the linear Kaiser model \cite{Kaiser:1987}, to combine with the halo model in previous studies, e.g., \cite{Seljak:2001}.
An alternative way to overcome the discrepancy between the halo and galaxy density fields is  to remove the effect of satellite kinematics from the observed galaxy distribution, known as halo density reconstruction \cite{Tegmark:2006, Reid:2009, Reid:2009a}.
We will not pursue this method here, but we note that it may be a useful alternative to the modeling developed here.

In this paper we present a theoretical model for the redshift-space power spectrum of galaxies using $N$-body simulations and halo PT. 
We use PT to model RSD for halos and add halo model inspired terms to model clustering effects that arise from satellites 
inside the halos. 
Motivated by the concept of the halo model, we decompose correlations of central and satellite galaxies in a galaxy sample into terms arising from galaxies within the same halo and those from separate halos, known as the 1-halo and 2-halo terms, respectively. We consider a simple model where the redshift-space density field in a 2-halo term is described by a Kaiser term (the simplest case being the linear Kaiser factor,  $1+f\mu^2/b$, where $\mu$ is the cosine of the angle between the line of sight and the wavevector $\vk$, $b$ is the bias parameter and $f=\ln{\delta}/\ln{a}$.) and RSD in a 1-halo term is described by well-known damping functions characterized by the nonlinear velocity dispersion parameter that depends on the host halo mass, $\sigma_v^2(M)$. We study theoretical models of the Kaiser terms using $N$-body simulations and PT.

This paper is organized as follows. 
In section \ref{sec:model}, we describe the decomposition of the observed density field and relate contributions from central and satellite galaxies to 1-halo and 2-halo terms described by a halo model. Section \ref{sec:sim} describes the $N$-body simulations and how we construct a mock galaxy sample. We present measurements of real-space and redshift-space power spectra in \ref{sec:analysis}.
In section \ref{sec:modeling}, we examine if it is possible to describe the redshift-space power spectrum using two models: one based on $N$-body simulations (\ref{sec:sim_model}) and another based on nonlinear PT (\ref{sec:df_model}). Our conclusions are given in section \ref{sec:conclusion}.

\section{Formalism of redshift-space power spectrum of galaxies}\label{sec:model}

In galaxy surveys, we observe two types of galaxies that contribute differently to a power spectrum measurement: 
central galaxies which can be considered to move together with the host halos (but, see \cite{Hikage:2012, Hikage:2013}) and satellite galaxies that mainly populate the most massive halos. 
Modeling the latter part is a nontrivial task, making it difficult to theoretically predict the statistics of a galaxy sample. 
Our goal here is to investigate the effects of satellite galaxies. Note that even the definition of the central galaxy is 
model-dependent for two reasons. One is that
it depends on the halo finder. Some halo finders tend to merge small halos into larger ones (e.g. friends of friends, FoF \cite{Davis:1985}), 
more than others (e.g. spherical overdensity). So what is a halo center for one halo may be a satellite for another. We will 
not address this issue here, and instead we will work with FoF halos only. 
Second reason is that the assignment of the halo center is also model dependent: center can be assigned to the most bound 
particle, or to the center of mass of all halo particles, among other choices. We will use the latter in this paper. 

One can always decompose the density field of galaxies in redshift space $\delta_g^S$ into 
contributions from central and satellite galaxies, respectively denoted as $\delta_ c^S$ and $\delta_s^S$. 
The superscript $S$ means a quantity defined in redshift space, while the superscript $R$ will denote the corresponding quantity in real space.
We can describe the decomposition in Fourier space as
\be
\delta_g^S(\vk) = (1-f_{s})\delta_c^S(\vk) + f_{s}\delta_s^S(\vk),
\ee
where $f_{s}=N_s/N_g=(1-N_c)/N_g$ is the satellite fraction, $N_c$ and $N_s$ are the number of central and satellite galaxies, respectively, and $N_g=N_c+N_s$ is the total number of galaxies.  
Note that the expression in redshift space can always be applied to the one in real space by looking at the tangential mode, $\delta_g^R(k)=\delta_g^S(k,\mu=0)$. 
The galaxy power spectrum is then written as the summation of
the spectra of central galaxies, satellite galaxies, and their cross-correlation, 
\bey
 P^S_{gg}(\vk)&=&(1-f_{s})^2 P^S_{cc}(\vk) \nn \\
 &+& 2f_{s}(1-f_{s}) P^S_{cs}(\vk) 
 +f_{s}^2 P^S_{ss}(\vk), \label{eq:pgg_pcc_pcs_pss}
 \eey
where $P_{XY}^S(\vk)(2\pi)^3\delta(\vk-\vk')\equiv\left\langle \delta_X^S(\vk)\delta_Y^{S*}(\vk') \right\rangle$ is a power spectrum of fields $X$ and $Y$ (auto spectrum if $X =Y$ and cross spectrum if $X\neq Y$).
Again, the corresponding real-space power spectrum can be obtained by $P^R_{XY}(k)=P^S_{XY}(k,\mu=0)$.
In a halo model approach, e.g., \cite{Seljak:2000, Peacock:2000, Ma:2000, Scoccimarro:2001, Cooray:2002}, the galaxy power spectrum can be decomposed into 1-halo and 2-halo terms. 
The halo model, originally developed to model the real-space power spectrum, was extended to redshift space by \cite{Seljak:2001, White:2001, Tinker:2007} (see also \cite{Hikage:2012}).
In order to decompose the observed power spectrum (equation \ref{eq:pgg_pcc_pcs_pss}) into contributions from 1-halo and 2-halo terms, we consider further divisions of central and satellite galaxies in the following subsections. 

\subsection{Decomposition of central galaxies}
Satellites and centrals live inside halos of a wide range of masses, and the corresponding bias terms, and FoG terms, can be 
very different. We decompose the central galaxies into 
two subsamples, those whose host halos do not and do contain satellite galaxies, respectively labeled as $c_A$ and $c_B$.  
In this case, the cross power spectrum between central and satellite galaxies can be written as 
\be
N_c P^S_{cs}(\vk)=N_{c_A}P^S_{c_As}(\vk)+N_{c_B}P^S_{c_Bs}(\vk),
\ee
where $N_{c}=N_{c_A}+N_{c_B}$. The sample $c_A$ consists of central galaxies that do not have any satellite galaxies in the same halo, so the contribution to the cross-correlation $P^S_{c_As}$ comes only from a 2-halo term. On the other hand, $P^S_{c_Bs}$ contains a 1-halo contribution, so we write it as $P^S_{c_Bs}=P^{S1h}_{c_Bs}+P^{S2h}_{c_Bs}$. A similar decomposition scheme was used by \cite{Zu:2008,Hikage:2013}.

\subsection{Decomposition of satellite galaxies}
Similarly, the satellite galaxy subsample can be decomposed into 
two subsamples. Satellite galaxies in halos with only a single satellite are denoted as sample $s_A$, and those in halos with at least one other satellite are denoted as sample $s_B$. 
Then, the auto-correlation of satellites $P^S_{ss}$ can be written as 
\bey
N_s^2 P^S_{ss}(\vk)=N_{s_A}^2P^S_{s_As_A}(\vk)&+&2N_{s_A}N_{s_B}P^S_{s_As_B}(\vk) \nn \\
&+&N_{s_B}^2P^S_{s_Bs_B}(\vk),
\eey
where $N_s=N_{s_A}+N_{s_B}$. 
Just like the case of $P^S_{c_As}$, the terms $P^S_{s_As_A}$ and $P^S_{s_As_B}$ only have contributions from 2-halo terms. 
$P^S_{s_Bs_B}$ includes a 1-halo contribution, so we write it as $P^S_{s_Bs_B}=P^{S1h}_{s_Bs_B}+P^{S2h}_{s_Bs_B}$. 

\subsection{Putting it all together}
Combining the terms outlined in the previous sections, the galaxy power spectrum in redshift space is
\be
P^S_{gg}(\vk)=P^{S1h}_{gg}(\vk) + P^{S2h}_{gg}(\vk), \label{eq:halo_model}
\ee
where the 2-halo and 1-halo terms are given by
\bey
P^{S2h}_{gg}(\vk) 
&=&(1-f_s)^2P^S_{cc}(\vk)+2f_s(1-f_s) \nonumber \\
& & \ \ \ \ \ \ \ \ \ \times\left( \frac{N_{c_A}}{N_c}P^S_{c_As}(\vk) + \frac{N_{c_B}}{N_c}P_{c_Bs}^{S2h}(\vk)\right) \nonumber \\
& +& f_s^2\left( \frac{N_{s_A}^2}{N_s^2}P_{s_As_A}^S(\vk)+\frac{2N_{s_A}N_{s_B}}{N_s^2}P_{s_As_B}^S(\vk) \right. \nonumber \\
& & \ \ \ \ \ \ \ \ \ + \left. \frac{N_{s_B}^2}{N_s^2}P_{s_Bs_B}^{S2h}(\vk)\right),\label{eq:full_2halo} \\
P^{S1h}_{gg}(\vk)& =&  2f_s(1-f_s)\frac{N_{c_B}}{N_c}P_{c_Bs}^{S1h}(\vk) \nn \\
& & \ \ \ \ \ \ \ \ \ + f_s^2\frac{N_{s_B}^2}{N_s^2}P_{s_Bs_B}^{S1h}(\vk). \label{eq:full_1halo}
\eey
Our goal is to compare the modeling of the individual terms to simulations. 
In our previous work \cite{Okumura:2012b, Vlah:2013}, we presented a theoretical modeling of the halo power spectrum in redshift space based on $N$-body simulations and perturbation theory, respectively. In section \ref{sec:modeling}, we examine if this scheme can be applied to the 2-halo terms of the galaxy power spectrum given in equation \ref{eq:full_2halo}.
Since we are primarily interested in modeling of RSD, we also present models where we use halo clustering from simulations.


\section{$N$-body simulations and mock galaxy samples}\label{sec:sim}
As in our previous work, e.g., \cite{Okumura:2012b}, we use a set of $N$-body simulations of the $\Lambda$CDM cosmology seeded with Gaussian initial conditions \cite{Desjacques:2009}. 
The primordial density field is generated using the matter transfer function by CAMB \cite{Lewis:2000}.
We adopt the standard $\Lambda$CDM model with $\Omega_m=1-\Omega_\Lambda=0.279$, $\Omega_b=0.0462$, 
$h = 0.7$, $n_s = 0.96$, $\sigma_8=0.807$ \cite{Komatsu:2009}.
We employ $1024^3$ particles of mass $m_p = 2.95\times 10^{11}h^{-1}M_\odot$ in 12 cubic boxes of a side $1600\himpc$. 
Dark matter halos are identified using the friends-of-friends algorithm \cite{Davis:1985} with a linking length equal to 0.17 times the mean particle separation.  We use all the halos with equal to or more than 20 particles, thus the minimum halo mass is $5.9\times 10^{12}h^{-1}M_\odot$.  
Because we consider the ``CMASS" galaxy sample from the Baryon Oscillation Spectroscopic Survey (BOSS) \cite{Schlegel:2009, Eisenstein:2011} as a target sample in this paper, we choose the output redshift of the simulations as $z=0.509$, which will be quoted as $z=0.5$  in what follows for simplicity.

\begin{table*}[bt!]
\caption{Properties of mock galaxy samples. $N_X$ is the number of objects in a given (sub)sample $X$, $\bar{M}$ is the average mass of the host halos, $\bar{n}_X=N_X/V$ is the number density, and $b_{1,X}$ is the large-scale bias determined using $P_{mX}^R(k)/P_{mm}^R(k)$ at $0.01\leq k \leq 0.04 \hmpci$.
}
\begin{center}
\begin{tabular}{ccccccc}
\hline\hline
Label & Galaxy/halo & $N_X$ & Fraction to & $\bar{M}$ &  $\bar{n}_X$ & \\
$X$ &     types     & ($\times10^4$) & total & $(10^{12}h^{-1}M_\odot)$ &  $(h^3{\rm Mpc}^{-3})$ & $b_{1,X}$  \\
\noalign{\hrule height 1pt}
$g$ & all galaxies & 125 & 1 & &   $3.03\times 10^{-4}$ &$2.17$   \\
$c$ & central galaxies& 109 & 0.877 & 26.25  & $2.67\times 10^{-4}$  &$2.02$  \\
$s$ & satellite galaxies& 15.3 & 0.123 & 106.8&  $3.75\times 10^{-5}$& $3.26$   \\
$c_A$ & centrals without satellite & 98.0 & 0.786 & 20.32  & $2.39\times 10^{-4}$&$1.91$  \\
$c_B$ & centrals with satellite(s) & 11.4 & 0.091 & 77.50  &  $2.77\times 10^{-5}$ &$2.92$  \\
$s_A$ & satellites without other satellite & 8.71 & 0.0698 & 58.61  & $2.13\times 10^{-5}$&$2.68$  \\
$s_B$ & satellites with other satellite(s)      & 6.63 & 0.0532 & 169.9  & $1.62\times 10^{-5}$&$4.00$  \\
bin2 & 2nd halo mass bin & 44.8 & & 28.99 & $1.09 \times 10^{-4}$ & 2.16 \\
bin3 & 3rd halo mass bin & 9.96 & & 85.37 & $2.43 \times 10^{-5}$ & 3.12 \\
 \hline \hline
\end{tabular}
\end{center}
\label{tab:gal_halo}
\end{table*}

To construct a mock galaxy catalog, we adopt a halo occupation distribution (HOD) model which populates dark matter halos with galaxies according to the halo mass, e.g., \cite{Cooray:2002, Zheng:2005}. 
Using the best fitting HOD parameters determined by \cite{White:2011} for the BOSS CMASS sample, galaxies are assigned to the halos at $z=0.5$. For halos which contain satellite galaxies, we randomly choose the same number of dark matter particles to represent the positions and velocities of the satellites. The fraction of satellite galaxies is determined to be $f_{s}=0.123$, consistent with the HOD modeling of \cite{White:2011}.
This method was applied in our previous work and good agreement with the observations has been confirmed for the correlation function \cite{Okumura:2012b} and mean pairwise infall momentum \cite{Okumura:2014}. 
The fraction of central galaxies that have satellites in the same halos relative to all the central galaxies is 
$N_{c_B}/N_{c}= 0.104$. 
Likewise, the fraction of satellite galaxies that have another satellite(s) inside the same halo relative to the total number of satellites is
$N_{s_B}/N_{s}= 0.432$. 
Table \ref{tab:gal_halo} summarizes the properties of the mock galaxy samples. 

In order to examine the effects of satellite galaxies on our statistics, we also analyze halo samples which have halo bias similar to the biases of our galaxy samples. We consider two halo subsamples used in our previous work \cite{Okumura:2012b}, denoted as ``bin2'' and ``bin3'', respectively. These halo samples have biases similar to those of the total galaxy sample and satellite galaxy sample considered in this work. The properties of the halo subsamples are also shown in table \ref{tab:gal_halo}.


\section{Numerical analysis}\label{sec:analysis}
Following our previous work \cite{Okumura:2012, Okumura:2012b}, we measure power spectra of given samples with a standard method. We compute the density field in real space or in redshift space on a grid of $1024^3$ cells using cloud-in-cell interpolation. When measuring the density field in redshift space, the positions of objects are distorted along the line of sight according to their peculiar velocities before they are assigned to the grid. We use a fast Fourier transform to measure the Fourier modes of the density field of sample $X$, $\delta_X(\vk)$, and then compute the power spectrum by multiplying the modes of the two fields (or squaring in the case of auto-correlation) and averaging over the modes within a wavenumber bin. To show the error of the mean for measured statistics, we divide the scatter among realizations by the square root of the number of the realizations, $1/\sqrt{12}$ in our case. 
For the redshift-space power spectrum we regard each direction along the three axes of simulation boxes as the line of sight; thus, the statistics are averaged over the three projections of all realizations for a total of 36 samples. The three measurements along different lines of sight are, however, not fully independent. To be conservative, we present the measured dispersion divided by $\sqrt{12}$ as the error of the mean even for the redshift-space spectra.


\subsection{Real-space power spectra}\label{sec:real_space_clustering}
\begin{figure}[bt] 
  \includegraphics[width=\linewidth]{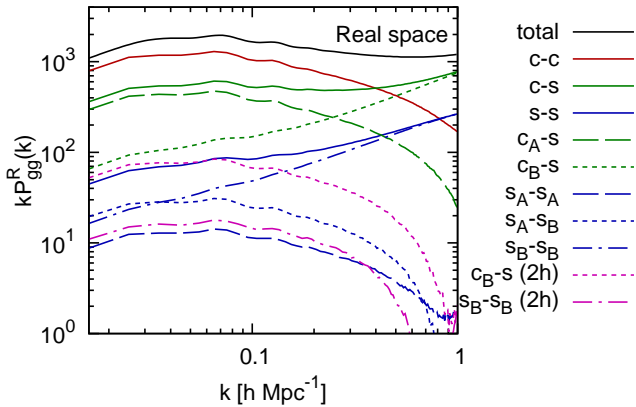}
\caption{Power spectrum of mock galaxy sample in real space $P^R_{gg}(k)$ (black). 
The red, green and blue solid lines are respectively the contributions from central-central ($P^R_{cc}$), central-satellite ($P^R_{cs}$), and satellite-satellite ($P^R_{ss}$) pairs to the full galaxy power spectrum. The central-satellite correlation can be further decomposed into the correlation between centrals that do not have a satellite in the same halo and satellites $P^R_{c_As}$ (green dashed) and that between centrals with satellite(s) and satellites $P^R_{c_Bs}$ (green dotted). Likewise, the satellite-satellite correlation can be decomposed into the auto-correlation of satellites that do not have other satellites inside the same halos $P^R_{s_As_A}$ (blue dashed), the auto-correlation of those that have other satellites inside the same halos $P_{s_Bs_B}^R$ (blue dot-dashed), and their cross-correlations $P_{s_As_B}^R$ (blue dotted). 
The 2-halo contributions of $P_{c_Bs}^R$ and $P_{s_Bs_B}^R$ are shown as the magenta dotted and dot-dashed lines, respectively.
}
\label{fig:lrg_cen_sat_real}
\end{figure}

Let us first consider the real-space galaxy clustering: the auto power spectrum of a given sample $P_{XX}^R(k)$ and the cross power spectrum of two different samples $P_{XY}^R(k)$. 
For the auto spectrum, we need to estimate and subtract shot noise from the measured spectrum $\widetilde{P}_{XX}^R$ (the tilde denotes a quantity directly measured from simulations), which is not a trivial task. Here, we assume a Poisson model, where the shot noise for the measured $P_{XX}^R$ is expressed as a constant,
\be \sigma_{n,X}^2=V/N_X=\bar{n}_X^{-1}, \label{eq:poisson_shot}\ee 
and we have
\be
P_{XX}^R(k)=\widetilde{P}_{XX}^R(k)-\sigma_{n,X}^2.
\ee
We show the real-space power spectra for our mock galaxy samples in figure \ref{fig:lrg_cen_sat_real}. The power spectrum measured for the full galaxy sample $P_{gg}^R$ is shown as the black line. The full galaxy power spectrum is decomposed into central-central $P_{cc}^R$, central-satellite $P_{cs}^R$, and satellite-satellite $P_{ss}^R$ correlations, as given in equation \ref{eq:pgg_pcc_pcs_pss}. Those three contributions are respectively shown as the red, green, and blue solid lines. 
While the auto spectrum of centrals $P^R_{cc}$ is dominant on large scales because of the low satellite fraction, the contributions from $P^R_{cs}$ and $P^R_{ss}$ become larger on small scales due to 1-halo terms. 

The contributions from the cross-correlations between satellites and centrals without ($P^R_{c_A s}$) and with ($P^R_{c_B s}$) a satellite inside the same halo, decomposed from $P_{cs}$, are respectively shown as the green dashed and dotted lines. As expected, the large amplitude of $P_{cs}^R$ at small scales is caused by $P^R_{c_B s}$, which includes a 1-halo term.  Similarly, the contributions from satellite-satellite correlations are decomposed as the blue dashed, dotted, and dot-dashed lines for $P^R_{s_As_A}$, $P^R_{s_As_B}$, and $P^R_{s_Bs_B}$, respectively. Once again, the small-scale power is dominated by $P^R_{s_Bs_B}$, due to the 1-halo contribution.

To eliminate the 1-halo contributions in $P_{c_Bs}$ and $P_{s_Bs_B}$, we consider a subsample of satellites whose positions are replaced by halo centers. In this case, the shot noise for the cross power between the centrals that have satellites in the same halo and the satellites is given by $\Sigma_{c_B}^2$ and
\be
P^{R2h}_{c_Bs}(k)=\widetilde{P}^{R2h}_{c_Bs}(k)-\Sigma_{c_B}^2, \label{eq:p_cbs_r2h}
\ee
where $\widetilde{P}^{R2h}_{c_Bs}$ expresses the spectrum measured from the satellite sample whose positions are replaced by halo centers, and
\be
\Sigma_{c_B}^2=V/N_{c_B}=1/\bar{n}_{c_B}. \label{eq:Sigma_cbs}
\ee
Likewise, the auto power spectrum of the satellites that have another satellite(s) in the same halo, whose positions are replaced by halo centers, is given by 
\be
P^{R2h}_{s_Bs_B}(k)=\widetilde{P}^{R2h}_{s_Bs_B}(k)-\sigma_{n,s_B}^2-\Sigma_{s_B}^2, \label{eq:p_sbsb_r2h}
\ee
where $\sigma_{n, s_B}^2$ is the normal Poisson shot noise (equation \ref{eq:poisson_shot}) and 
\be
\Sigma_{s_B}^2=V\frac{\sum_i^{N_c}N_{s,i}(N_{s,i} - 1) }{ \left( \sum_i^{N_c} N_{s,i} \right)^2}, \label{eq:Sigma_sbsb}
\ee
where $N_{s,i}$ is the number of satellites in the $i^\mathrm{th}$ halo, and the sum is over all halos that host more than one satellite galaxy. In our case, $\Sigma_{s_B}^2=4.42\times 10^{-5}V-\sigma_{n, s_B}^2=2.92\times 10^{-5}V=1.19\times 10^{5} \ (\hmpc)^3$.
These measurements (equations \ref{eq:p_cbs_r2h} and \ref{eq:p_sbsb_r2h}) are shown as the dotted and dash-dotted magenta lines in figure \ref{fig:lrg_cen_sat_real}, respectively. One can see that the small-scale power caused by the 1-halo terms is well suppressed, and the shapes of the power spectra $P^{R2h}_{c_Bs}$ and $P^{R2h}_{s_Bs_B}$ are similar to the other 2-halo spectra.



\subsection{Galaxy biasing}\label{sec:bias}
\begin{figure}[bt] 
  \includegraphics[width=\linewidth]{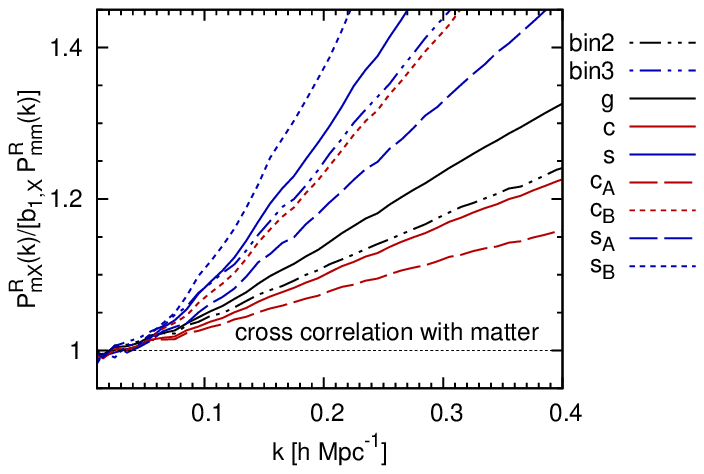}
  \includegraphics[width=\linewidth]{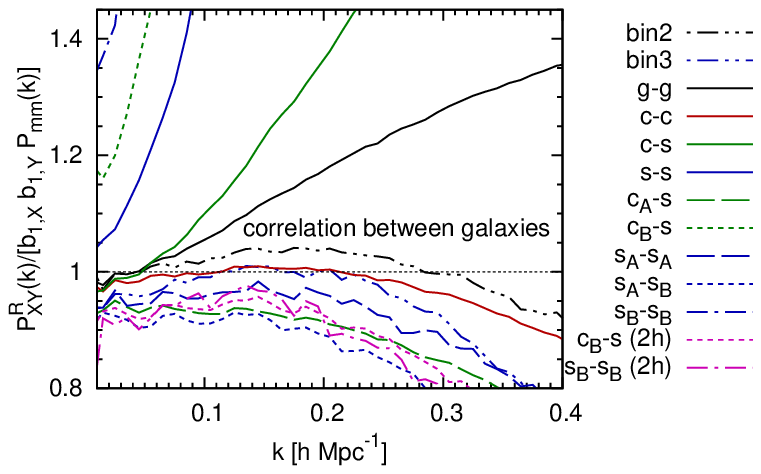}  
\caption{({\it top}) The cross-correlation $P_{mX}^R$ normalized by $b_{1,X} P_{mm}^R$, where $X$ denotes
samples described in the panel and $b_{1,X}$ is the linear bias determined in the large-scale limit. ({\it bottom}) The normalized auto spectrum $P_{XX}^R/b_{1,X}^2P_{mm}^R$ and 
cross spectrum of the samples $X$ and $Y$, $P_{XY}^R/b_{1,X}b_{1,Y}P_{mm}^R$.
The magenta dotted and dot-dashed lines are respectively the results for $P_{c_B s}^R$ and $P_{s_Bs_B}^R$ where the 1-halo contributions are eliminated. 
}
\label{fig:bias}
\end{figure}

The bias for a given galaxy sample $X$, $b_X(k)$, can be defined as
\be
b_X(k)=\frac{P_{mX}^R(k)}{P_{mm}^R(k)},
\ee
where $P_{mm}^R$ is the auto spectrum of dark matter, and $P_{mX}^R$ is the cross power spectrum between the sample $X$ and dark matter. The bias defined using the cross power spectrum is not affected by the shot noise. The value of the linear bias parameter, $b_{1,X}$, is determined by minimizing $\chi^2$ statistics over the wavenumber range $0.01\leq k\leq 0.04 \hmpci$. 
The top panel of figure \ref{fig:bias} shows the galaxy biasing normalized by the linear bias, $b_X(k)/b_{1,X}=P_{mX}^R(k)/\left[P_{mm}^R(k)b_{1,X}\right]$. In the large-scale limit, the normalized bias for all the samples approaches unity. 
The values of $b_{1,X}$ in the large-scale limit are summarized in table \ref{tab:gal_halo}.
As we have seen for halos in \cite{Okumura:2012b}, the galaxy bias deviates from a constant at increasingly larger scales for a more biased sample. For a comparison, we show the results for the two halo mass bin samples measured in \cite{Okumura:2012b} as $[P_{00}^{mh}(k)]^{1/2}$, each of which has a bias value similar to the whole galaxy sample (``bin2'' halos) and the satellite sample (``bin3'' halos). 
We see that they are similar, although not identical, possibly a consequence of having a different halo mass distribution. 
However, some of the difference appears to be simply an error in the overall bias due to noise in the bias 
determination at low $k$. 

We next look at the nonlinearity of the auto spectrum of a given galaxy sample $X$, $P_{XX}^R(k)$, and the cross spectrum between samples $X$ and $Y$, $P_{XY}^R(k)$, using the bias parameters determined above. The bottom panel of figure \ref{fig:bias} shows the auto spectrum normalized as $P_{XX}^R(k)/\left[P_{mm}^R(k)b_{1,X}^2\right]$ and the cross power as $P_{XY}^R(k)/\left[P_{mm}^R(k)b_{1,X}b_{1,Y}\right]$, where $b_{1,X}$ is again the large-scale limit of $b_X(k)$ determined from the cross-power spectrum with matter $P_{mX}^R$. The black solid line is the result for the whole galaxy sample $P_{gg}^R$. The red and blue lines show the results for the auto spectra of centrals and satellites, respectively, while the green line shows the result for the cross spectrum between them. The shot noise for the spectra that have 1-halo contributions is known to deviate from unity in the large-scale limit, e.g., \cite{Yu:1969, Peebles:1980, Seljak:2009a, Baldauf:2013}. 
Although $P_{cs}^R$ also contains a 1-halo contribution (that is, $P_{c_Bs}^{R1h}$), it is a minor effect of order $\sim$10\%, (see section \ref{sec:sim}), and thus $P_{cs}^R/\left[P_{mm}^R b_{1,c}b_{1,s}\right]$ is consistent with unity in the $k\to 0$ limit. However, the deviation for $P_{c_Bs}^R$ from unity is more prominent at large scales. Unlike $P_{cs}^R$, one can see the clear deviation for $P_{ss}^R$ because the 1-halo effect is $\sim$43\%. The normalized power spectra that have only 2-halo contributions are well described by constants in the $k\to 0$ limit. Even the results for the spectra $P_{c_Bs}^R$ and $P_{s_Bs_B}^R$, after eliminating 1-halo contributions, become constant. These results confirm that the normalized spectra with 2-halo terms have the same power spectrum shape. 
We again see some small 
differences between the halo sample corresponding to the same bias as the centrals, and the centrals. Some of 
this is due to an overall error in bias determination. 


\subsection{Halo satellite radius}

In this subsection we examine if the effect of the halo profile on 1-halo and 2-halo terms can be modeled by the leading order correction which is proportional to $k^2R_X^2$ where $R_X$ is the typical radius for a given galaxy sample $X$ inside halos. This model will be tested with the simulation measurements.

We can write the measured power spectra that have both 1-halo and 2-halo contributions, $\widetilde{P}^{R}_{XY}$ where $XY=\{c_Bs, \ s_Bs_B\}$, as 
\bey
\widetilde{P}^{R}_{c_Bs}(k)&=& P^{R1h}_{c_Bs}(k)+P^{R2h}_{c_Bs}(k), \label{eq:shot1_cbs}\\
\widetilde{P}^{R}_{s_Bs_B}(k)&=& P^{R1h}_{s_Bs_B}(k)+P^{R2h}_{s_Bs_B}(k)+\sigma_{n,s_B}^2. \label{eq:shot1_sbsb}
\eey
The 2-halo terms $P^{R2h}_{XY}$ are related to the measured spectra $\widetilde{P}^{R2h}_{XY}$ through equations \ref{eq:p_cbs_r2h} and \ref{eq:p_sbsb_r2h}. 
In the $k\to 0$ limit, the 1-halo terms $P^{R1h}_{XY}$ are a constant and behave as white noise, $P^{R1h}_{c_Bs}=\Sigma_{c_B}^2$
and $P^{R1h}_{s_Bs_B}=\Sigma_{s_B}^2$ (see, e.g., \cite{Baldauf:2013}). 
At high $k$, the halo density profile damps both the 1-halo and 2-halo terms  \cite{Seljak:2000,Mohammed:2014}, and we consider the leading-order $k^2R^2$ corrections for both terms, where $R$ is the typical halo radius. Thus, the 1-halo term deviates from a constant as 
\be
P^{R1h}_{XY}(k)=\Sigma_{Y}^2\left(1-k^2 [R_X^2+R_Y^2] \right). \label{eq:excl_1h}
\ee
Note that for $P_{c_Bs}$, $R_{c_B}\ll R_{s}$, so we consider only $R_s$ and set $R_{c_B}=0$.
Similarly, when including the leading-order profile correction, the 2-halo terms become
\bey
P^{R2h}_{c_Bs}(k) = \widetilde{P}^{R2h}_{c_Bs}(k) &-& \Sigma_{c_B}^2  \nn \\
												&-& b_{1,c_B}b_{1,s}P_{\rm lin}(k)k^2 R_{s}^2,  \label{eq:excl_2h_cbs} \\
P^{R2h}_{s_Bs_B}(k) = \widetilde{P}^{R2h}_{s_Bs_B}(k) &-& \sigma_{n,s_B}^2 \nn \\
											&-& \Sigma_{s_B}-2b_{1,s_B}^2P_{\rm lin}(k)k^2 R_{s_B}^2. \label{eq:excl_2h_sbsb}
\eey
By inserting equation (\ref{eq:excl_1h}) into equations (\ref{eq:shot1_cbs}) and (\ref{eq:excl_2h_cbs}) for $P^R_{cBs}$ and (\ref{eq:shot1_sbsb}) and (\ref{eq:excl_2h_sbsb}) for $P^R_{s_Bs_B}$, we obtain
\bey
\widetilde{P}^{R2h}_{c_BS}-\widetilde{P}^R_{c_Bs}&=& \left[\Sigma_{c_B}^2  + b_{1,c_B}b_{1,s}P_{\rm lin}\right]k^2 R_{s}^2, \label{eq:excl_formula1} \\
\widetilde{P}^{R2h}_{s_Bs_B}-\widetilde{P}^R_{s_Bs_B}&=&2\left[\Sigma_{s_B}^2+b_{1,s_B}^2P_{\rm lin}\right]k^2 R_{s_B}^2. \label{eq:excl_formula2}
\eey
They are the results of the model using the leading-order $k^2R^2$ corrections. 

\begin{figure}[bt] 
  \includegraphics[width=\linewidth]{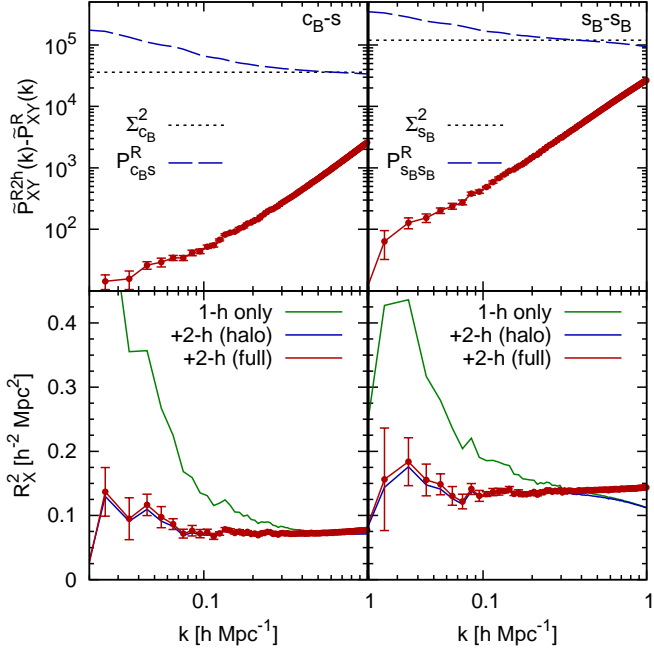}
\caption{{\it Top left panel}: The difference between the power spectrum of centrals (inside halos with satellites, $c_B$) 
and satellites, and the power spectrum of same centrals with satellites replaced by halo centers, 
$\widetilde{P}_{c_Bs}^{R2h}-\widetilde{P}_{c_Bs}^{R}$ (red). The blue dashed line shows $P_{c_Bs}^R(=\widetilde{P}_{c_Bs}^R)$. The horizontal dotted line is the expected 1-halo term at the large-scale limit, $P_{c_Bs}^{R1h}=\Sigma_{c_B}^2$.  {\it Top right panel}: Same as the top left panel but for the power spectrum of satellites that have one or more satellite in the same halo, $\widetilde{P}_{s_Bs_B}^{R2h}-\widetilde{P}_{s_Bs_B}^{R}$. The blue dashed line shows $P_{s_Bs_B}^R=\widetilde{P}_{s_Bs_B}^R-\sigma_{n,s_B}^2$ and the horizontal line $P_{s_Bs_B}^{R1h}=\Sigma_{s_B}^2$. 
{\it Bottom left panel}: Typical satellite radius $R_{s}^2$ as a function of $k$ computed from equation (\ref{eq:excl_formula1}) with (blue) and without (green) the 2-halo correction, and from equation (\ref{eq:excl_full1}) (red). For clarity error bars are shown only for the latter result.
{\it Bottom right panel}: Same as the bottom left panel but for the auto spectrum, thus the formula is given by equations (\ref{eq:excl_formula2}) and (\ref{eq:excl_full2}).
}
\label{fig:diff_1h_2h}
\end{figure}

\begin{figure*}[bt] 
  \centering
  \includegraphics[width=.88\linewidth]{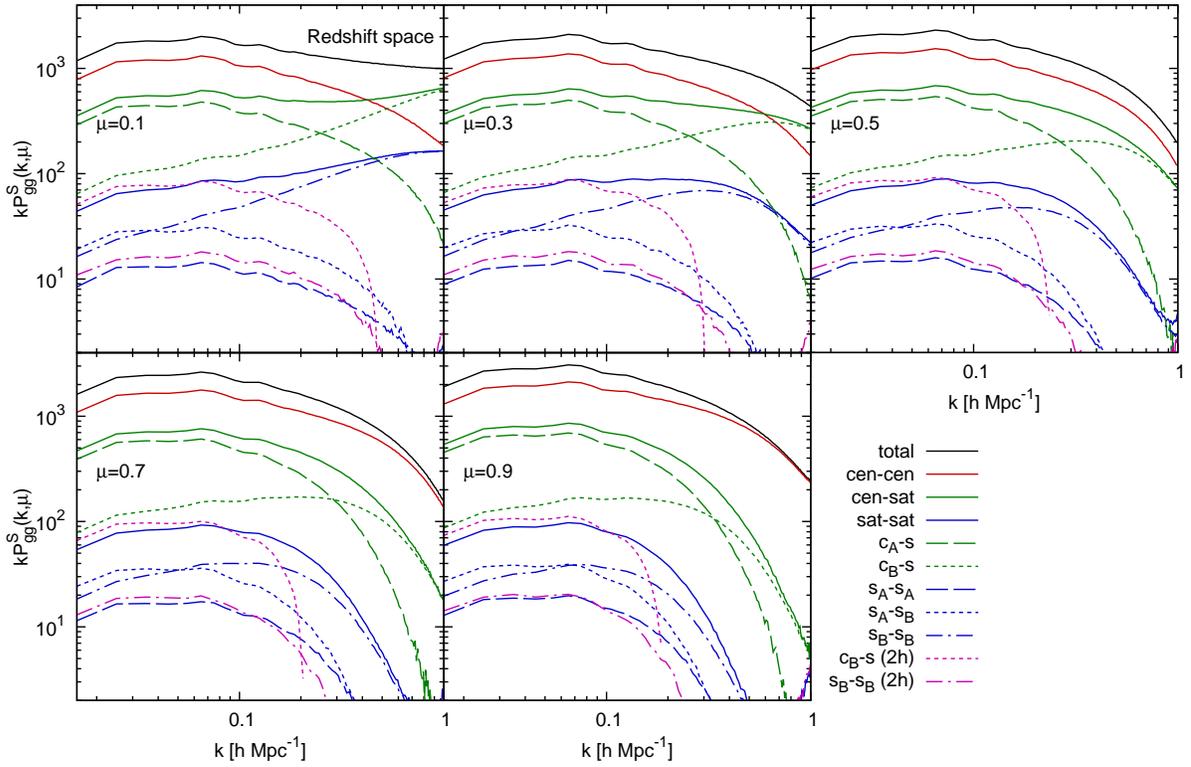}
\caption{2D Power spectrum of mock galaxy sample $P^S_{gg}(k,\mu)$ denoted as the black points with lines and the contributions from central and 
satellite galaxies to it. The width of $\mu$ bin is 0.2 centered around the values shown in each panel.
The meaning of the color and type of each line is  the same as that in figure \ref{fig:lrg_cen_sat_real}.
}
\label{fig:lrgmu_cen_sat}
\end{figure*}

Note that the terms inside the brackets of the right hand sides of equations \ref{eq:excl_formula1} and \ref{eq:excl_formula2} are 
simply a halo model expression of the measured spectra, so we can instead write
\bey
\widetilde{P}^{R2h}_{c_Bs}-\widetilde{P}^R_{c_Bs}        &=& \widetilde{P}^{R}_{c_Bs} k^2 R_{s}^2  , \label{eq:excl_full1} \\ 
\widetilde{P}^{R2h}_{s_Bs_B}-\widetilde{P}^R_{s_Bs_B}&=& 2 \left[\widetilde{P}^{R}_{s_Bs_B}-\sigma_{n,s_B}^2 \right] k^2 R_{s_B}^2.\label{eq:excl_full2} 
\eey

In the top left panel of figure \ref{fig:diff_1h_2h}, we show the result for the cross power spectrum between centrals that have one or more satellites and satellites, $\widetilde{P}^{R2h}_{c_B s}-\widetilde{P}^{R}_{c_B s}$. The horizontal solid line is the expected shot noise $\Sigma_{c_B}^2$. The blue dashed line shows the full power spectrum, $P_{c_Bs}^R (=\widetilde{P}_{c_B}^R)$. The bottom left panel shows the square of the halo radius, $R_{s}^2$, for our two models described above. 
The green lines are the result of equation \ref{eq:excl_formula1} when the profile correction to the 2-halo term (equation \ref{eq:excl_2h_cbs}) is ignored, $R_s^2=\left(\widetilde{P}_{c_B s}^{R2h}-\widetilde{P}_{c_B s}^{R}\right) / k^2\Sigma_{c_B}^2$. 
The rise of $R_s$ at large scales is caused by the absence of the correction. 
Including the 2-halo correction, presented as the blue line, makes the lines flat.
The result of the full model (equation \ref{eq:excl_full1}) is presented as the red line. The error bars are  
shown only for this result for clarity. 
The right panels are the same as the left panels but for the auto power spectrum of the satellites that have at least one other satellite in the same halo, $\widetilde{P}^{R2h}_{s_B s_B}-\widetilde{P}^{R}_{s_B s_B}$. 
The right bottom panel of figure \ref{fig:diff_1h_2h} shows our models for $R_{s_B}$. We find $R_{s_B}>R_{s}$ as expected, and also our models give nearly a constant value of $R_{s_B}$. 
The typical satellite radius is about $0.3$Mpc/h for central-satellite pairs,and slightly larger for satellite-satellite pairs, 
as expected since the latter are in larger halos. 

Our models with $R_s$ can correct the effects of the halo density profile,  and the parameters can be tied to the typical extent 
of satellites inside the halos. However, the effect on the full galaxy sample up to $k=0.5 \hmpci$
is small: the effects are of order 3\% at $k \sim 0.5\hmpci$, for central-satellite pairs, and of order 10\% for satellite-satellite 
pairs, but since the first one is downweighted by the satellite fraction, and the second one by satellite fraction squared, for 
a typical value of the satellite fraction of 10\% the overall effect is less than 1\%. 
Thus in the following analysis we do not include the profile correction in the modeling. 

\begin{figure*}[tbh] 
  \centering
  \includegraphics[width=.88\linewidth]{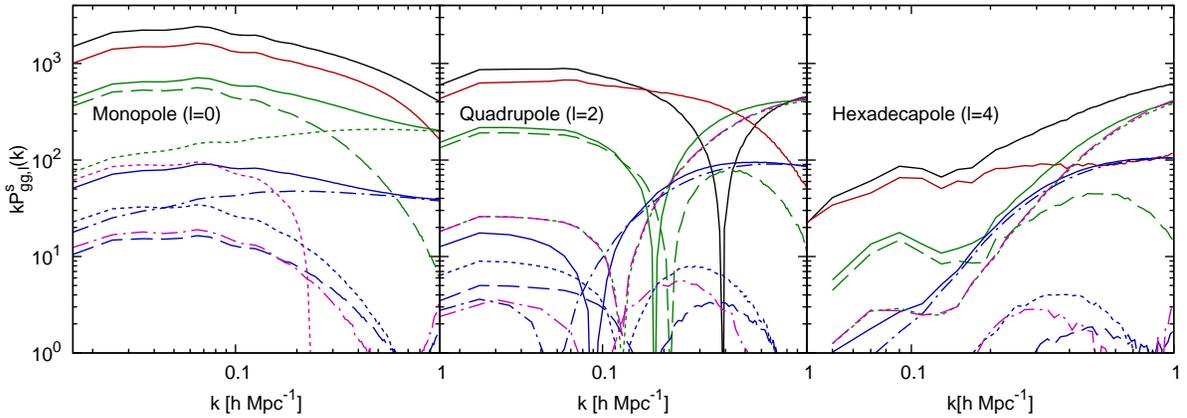}
\caption{
Multipole moments of galaxy power spectrum ($P_{gg,l}^S$), the monopole (left), quadrupole (middle) and hexadecapole (right). The meaning of the color and style of each line is the same as in figures \ref{fig:lrg_cen_sat_real} and \ref{fig:lrgmu_cen_sat}. The sign of the quadrupoles changes from positive to negative at small scales except for $P_{cc,2}^S$.
}
\label{fig:lrg_cen_sat}
\end{figure*}

\subsection{Redshift-space power spectra}\label{sec:redshift_space_clustering}
We present next the measurements of redshift-space power spectra, $P^S_{XX}$ and $P^S_{XY}$, where $X$ and $Y$ are the given samples, ($g$, $c$, $s$, and so on). As in the case of the real-space power spectra, the effect of shot noise for the auto power spectrum in redshift space is assumed to be Poisson, such that $P^S_{XX}(k,\mu)=\widetilde{P}^S_{XX}(k,\mu)-\sigma_{n,X}^2$.
In figure \ref{fig:lrgmu_cen_sat}, we show the redshift-space 2D power spectrum of the mock galaxy sample,
$P^S_{gg}(k,\mu)$, for the 5 different $\mu$ bins as the black solid lines. The red, blue, and green curves respectively show the 
contributions of the auto-correlations of central galaxies, satellite galaxies, and their cross-correlations to the full spectrum. 
Because $P^S(k,\mu=0) = P^R(k)$, all the spectra $P^S(k,\mu=0.1)$ shown in the top left panel are very similar to the results presented in figure \ref{fig:lrg_cen_sat_real}. 
At higher $\mu$, the spectra $P^S_{cs}$ and $P^S_{ss}$ are strongly suppressed at small scales because of the nonlinear velocity dispersion including the FoG effect, which also leads to the suppression of the total galaxy spectrum $P^S_{gg}$. 
The decomposed cross-power spectra between centrals and satellites, $P_{c_As}^S$ and $P_{c_Bs}^S$ are presented as the green dashed and dotted lines, respectively. 
The 1-halo term of $P_{c_Bs}^S$ dominates $P_{cs}^S$ at small scales, and one can see that the suppression of the amplitude of $P_{cs}^S$ for large $\mu$ is caused by the same term. Likewise, the decomposed auto spectra of satellites, $P_{s_As_A}$, $P_{s_As_B}$, $P_{s_Bs_B}$, are shown as the blue dashed, dotted and dot-dashed lines, and $P_{s_Bs_B}^S$, which has 1-halo contributions, becomes dominant at small scales. 

As we did in previous sections for real space, the 2-halo contributions of the two terms that also contain 1-halo contributions, $P_{c_Bs}^{S2h}$ and $P_{s_Bs_B}^{S2h}$, can be obtained by replacing satellite positions by halo centers. Here, we keep the  velocity of one of the satellites in the same halo chosen randomly. As in the case for real space (equations \ref{eq:p_cbs_r2h} and \ref{eq:p_sbsb_r2h}), we have
\bey
P^{S2h}_{c_Bs}(k,\mu)&=&\widetilde{P}^{S2h}_{c_Bs}(k,\mu)-\Sigma_{c_B}^2, \label{eq:p_cbs_s2h} \\ 
P^{S2h}_{s_Bs_B}(k,\mu)&=&\widetilde{P}^{S2h}_{s_Bs_B}(k,\mu)-\sigma_{n,s_B}^2-\Sigma_{s_B}^2, \label{eq:p_sbsb_s2h} 
\eey
where $\Sigma_{c_B}$ and $\Sigma_{s_B}$ are given in equations (\ref{eq:Sigma_cbs}) and (\ref{eq:Sigma_sbsb}).
They are shown in figure \ref{fig:lrgmu_cen_sat} as the dotted and dot-dashed magenta lines, respectively. 

When analyzing real data, the multipole moments of the redshift-space power spectrum are often used to reduce the degrees of freedom, thus simplifying the analysis. The multipoles are described using Legendre polynomials ${\cal P}_l(\mu)$ as 
\be
P^S_l(k)=(2l+1)\int^1_0 P^S(k,\mu){\cal P}_l(\mu) d\mu. 
\ee
Figure \ref{fig:lrg_cen_sat} presents the redshift-space multipoles: from the left, the monopole ($l=0$), quadrupole ($l=2$), and hexadecapole ($l=4$). Similar plots are shown by \cite{Hikage:2014}.
In the following section we will present results for the full 2D spectrum $P^S(k,\mu)$ rather than the multipoles in order to more closely examine any small deviations of our theoretical modeling from $N$-body measurements.


\section{Modeling the galaxy redshift-space power spectrum} \label{sec:modeling}

The galaxy power spectrum in redshift space in our formalism is given by equation \ref{eq:halo_model}, and in this section, we present the prediction for each term in equations \ref{eq:full_2halo} and \ref{eq:full_1halo}.
 We consider two models for the redshift-space power spectrum of galaxies: one based on $N$-body simulations (section \ref{sec:sim_model}) and another based on nonlinear perturbation theory (section \ref{sec:df_model}). 

Satellite galaxies have large, nonlinear velocities inside their host halos which suppress the clustering amplitude relative to linear theory at small scales, a phenomenon known as the FoG effect \cite{Jackson:1972}.
In previous studies, the effect has been modeled with a single damping factor $G(k\mu;\sigma_v)$, with $\sigma_{v}$ corresponding to the velocity dispersion of a given system, and the redshift-space power spectrum of galaxies modeled as 
 $P_{gg}^S(k,\mu)=G^2(k\mu;\sigma_v)P_{hh}^S(k,\mu)$, where $P_{hh}^S$ is the halo power spectrum, e.g., \cite{Peacock:1994, Park:1994, Scoccimarro:2004,Nishimichi:2011,Sato:2011}. 
 
Only the density field of satellite galaxies is affected by the nonlinear velocity dispersion.
Considering this fact, the four power spectra that have only 2-halo contributions, namely $P_{XY}^S$ where $XY=\{cc,c_As ,s_As_A, s_As_B\}$ (see equation \ref{eq:full_2halo}), are given by 
\bey
 	P_{cc}^S(k,\mu) &=& P_{cc,h}^S(k,\mu) \label{eq:p_cc} \\
 	P_{c_As}^S(k,\mu) &=&G(k\mu;\sigma_{v,s}) P_{c_As,h}^S(k,\mu),  \label{eq:p_cas} \\
 	P_{s_As_A}^S(k,\mu) &=& G^2(k\mu;\sigma_{v,s_A})P_{s_As_A,h}^S(k,\mu),  \label{eq:p_sasa}\\
 	P_{s_As_B}^S(k,\mu) &=& G(k\mu;\sigma_{v,s_A})G(k\mu;\sigma_{v,s_B})\nonumber \\
		&&\times P_{s_As_B,h}^S(k,\mu),  \label{eq:p_sasb} 
\eey
where $P^S_{XX,h}$ represents the auto power spectrum of halos in which the galaxies $X$ reside, and 
$P^S_{XY,h}$ the cross spectrum of halos in which galaxies $X$ and $Y$ reside. 
Note that the halo spectrum $P^S_{XX,h}$ or $P^S_{XY,h}$ needs to be distinguished from $P^{S2h}_{XX}$ or $P^{S2h}_{XY}$ presented in section \ref{sec:redshift_space_clustering} because the latter is the 2-halo term of the galaxy power spectrum, thus it is affected by the nonlinear velocity dispersion effect. 
For the Poisson shot noise model, we have $P_{XY,h}^S=\widetilde{P}_{XY,h}^S$ and $P_{XX,h}^S=\widetilde{P}_{XX,h}^S-\sigma_{n,X}^2$ for these four halo spectra. Under the assumption of linear perturbation theory, the spectrum $P_{XY,h}^S$ converges to the linear RSD power spectrum originally proposed by \cite{Kaiser:1987}, $P^S_{XY,h}(k,\mu)=(b_{1,X}+f\mu^2)(b_{1,Y}+f\mu^2)P_{\rm lin}(k)$, where $P_{\rm lin}$ is the linear power spectrum of underlying dark matter in real space.

The remaining two terms are the cross power spectrum between the centrals that have satellite(s) in the same halo and satellite galaxies, $P_{c_Bs}^S$, and the auto spectrum of satellites that have at least one other satellite in the same halo, $P_{s_Bs_B}^S$. 
As we have seen in section \ref{sec:model}, these spectra have both 1-halo and 2-halo contributions, and the shot noise deviates from a constant due to the scale-dependence of the 1-halo terms (although we ignore this effect in the full model because of low satellite 
fraction). 
Consequently, modeling these spectra is not as straightforward as for those terms that contain only 2-halo contributions. 

Nevertheless, we can follow the same procedure for the spectra that have 1-halo contributions. 
The cross spectrum of the centrals that contain satellites within the same halo and the satellites, $P_{c_Bs}^S(=\widetilde{P}_{c_Bs}^S)$, can be modeled as 
\bey
P_{c_B s}^S(k,\mu)
&=& G(k\mu;\sigma_{v,s}) \left[\widetilde{P}_{c_Bs,h}^S(k,\mu) - \Sigma_{c_B}^2\right] \nn \\
&&+G(k\mu;\sigma_{v,s})\Sigma_{c_B}^2 \nn \\
&=&
G(k\mu;\sigma_{v,s}) \widetilde{P}_{c_Bs,h}^S(k,\mu) \label{eq:p_cbs}
\eey
where the shot noise term $\Sigma_{c_B}^2$ is given by equation \ref{eq:Sigma_cbs}. 
Likewise, the auto power spectrum of the satellites that have another satellite(s) in the same halos, $P_{s_Bs_B}^S$, can be described as 
\bey
P_{s_B s_B}^S(k,\mu) &=& 
\widetilde{P}_{s_B s_B}^S(k,\mu) -\sigma_{n,s_B}^2 \nn \\ 
&=& G_{2h}^2(k\mu;\sigma_{v,s_B})  \nn \\
&& \times \left[\widetilde{P}_{s_Bs_B,h}^S(k,\mu) -\Sigma_{s_B}^2- \sigma_{n,s_B}^2 \right], \nn \\
&+& G_{1h}^2(k\mu;\sigma_{v,s_B}) \Sigma_{s_B}^2  \label{eq:p_sBsB}
\eey
where $\Sigma_{s_B}$ is given by (\ref{eq:Sigma_sbsb}). 
$G_{1h}$ and $G_{2h}$ are the damping factors due to the FoG effect on 1-halo and 2-halo terms, respectively. 
In the general case with a wide halo mass distribution they would be different because 1 and 2 halo terms 
have different weightings as a function of the halo mass, while for a narrow halo mass distribution 
we expect the same damping factor for the two, $G_{1h}=G_{2h}\equiv G$. Our goal is to divide up the correlations 
into individual contributions from components that have a relatively narrow mass distribution (centrals with and 
without satellites, satellites with other satellite pairs...), and we are thus trading the simplicity of the 
modeling of individual terms with narrow halo mass distribution with the complexity of having several terms that we need to model. 
Thus we have 
\be
P_{s_B s_B}^S(k,\mu)  
= G^2(k\mu;\sigma_{v,s_Bs_B}) \left[\widetilde{P}_{s_Bs_B,h}^S(k,\mu) -\sigma_{n,s_B}^2 \right]. \label{eq:p_sbsb2}
\ee
Equations \ref{eq:p_cc}--\ref{eq:p_sasb}, \ref{eq:p_cbs} and \ref{eq:p_sbsb2} are what we will predict with two ways in the following subsections. 
Note that even though our model splits FoG effects 
into 1 and 2 halo terms, we could have also modeled it simply by multiplying FoG terms on 
the total power spectrum combining the two, in analogy of halo profile effects, where equations \ref{eq:excl_formula1}-\ref{eq:excl_formula2}
are equivalent 
to equations \ref{eq:excl_full1}-\ref{eq:excl_full2}. 

Gaussian and exponential distributions are usually considered for pairwise velocity dispersion in configuration space, which correspond to the Gaussian and Lorentzian functions in Fourier space, respectively \cite{Peacock:1994, Park:1994, Percival:2009},
\bey
G(k\mu;\sigma_v)=\left\{ 
\begin{array}{ll}
e^{-k^2\mu^2\sigma_v^2/2} & {\rm Gaussian}, \\
\left( 1+ k^2\mu^2\sigma_v^2/2 \right)^{-2} & {\rm Lorentzian}.
\end{array}\right. \label{eq:fog_phenom}
\eey
Note that the Lorentzian function has a form modified from the commonly-used form in the literature. There is no formal way to 
derive these, so we simply adopt whichever works best. 
See \cite{Okumura:2012} for a more generalized functional form for the nonlinear velocity dispersion effect that approaches the Gaussian and Lorentzian forms as the two limit cases.  
Under the assumption that satellites follow the virialized motions inside the 
halos, the velocity dispersion of the satellite galaxies about the center of mass of the host halo can be described as a function of halo mass, e.g., \cite{Cooray:2002}, 
\be
\sigma_v^2(M)=\sigma_{v,0}^2 \left(\frac{M}{10^{13}M_\odot/h}\right)^{2/3}, \label{eq:vel_disp_model}
\ee
where the normalization $\sigma_{v,0}$ depends on both the assumed RSD model and the functional form of the damping term (equation \ref{eq:fog_phenom}). 
The suppression of the power at small scales due to the damping factor is different for different $P^S_{XY}$ terms because the galaxies in each subsample reside in halos with different mass. However, for each RSD model of the galaxy power spectrum $P^S_{gg}$, there is only one free parameter for the velocity dispersion, $\sigma_{v,0}$. In practice of course even this assumption is not valid to some
extent: one can have radial profile of satellites to vary with the halo mass in a way that does not scale with the virial radius, 
for example. We will ignore these considerations. 

\begin{figure*}[bt] 
  \includegraphics[width=\linewidth]{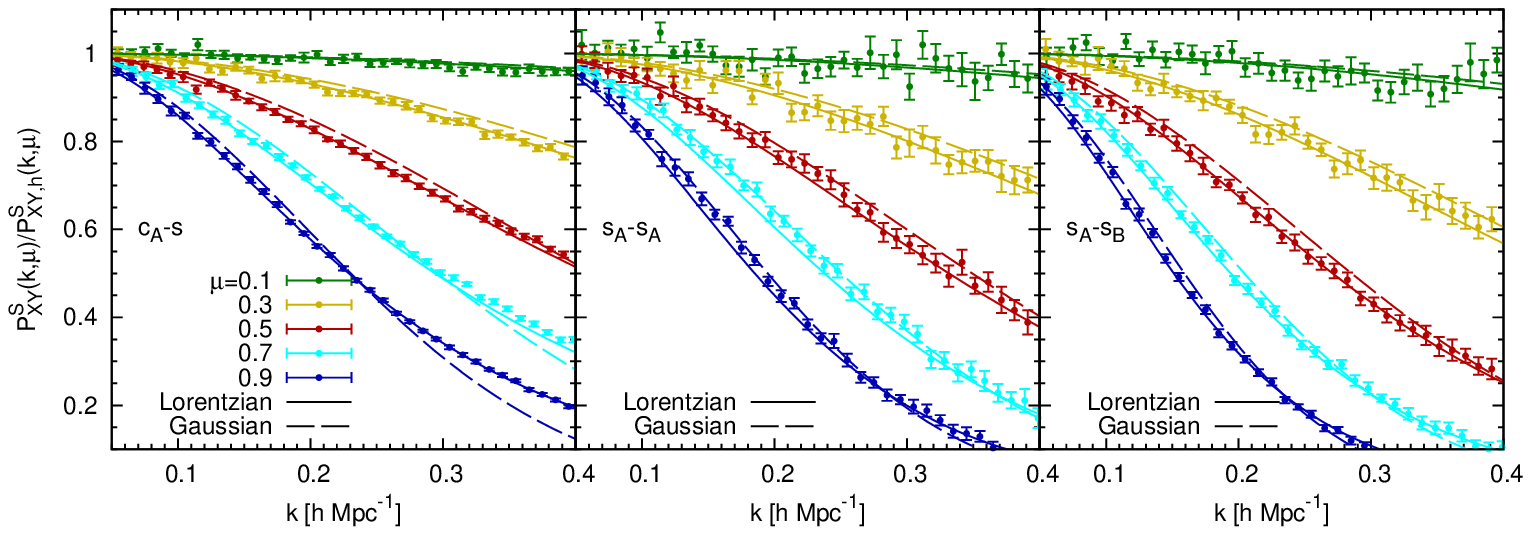}
  \includegraphics[width=\linewidth]{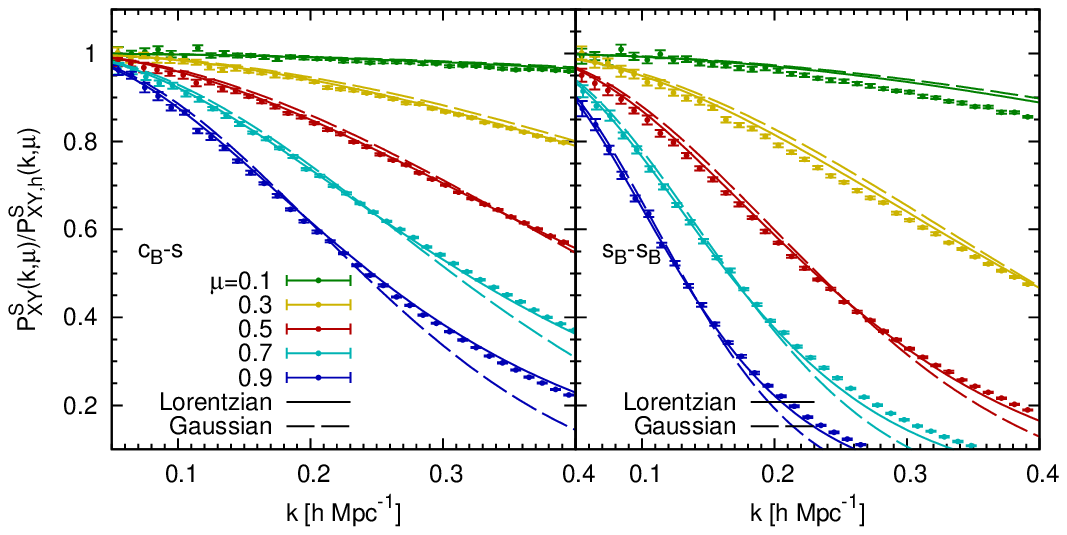}
\caption{The ratio of the redshift-space power spectra for various galaxy samples to the corresponding halo spectra.
The halo spectra are computed by replacing satellite positions and velocities by those of halos. The top panels show
spectra that contain only 2-halo contributions: the correlation between centrals without satellites and satellites (left), the auto-correlation of satellites that have no other satellites in the same halo (middle), and the cross-correlation between satellites with and without other satellites in the same halo (right). 
The bottom panels show the spectra which have both 1-halo and 2-halo contributions: the cross-correlation between the centrals that have satellite(s) in the same halo and satellite galaxies (left) and the auto-correlation of satellites that have at least one other satellite in the same halo (right).
The dashed and solid lines are respectively the predictions for the Gaussian and Lorentzian damping functions due to the nonlinear velocity dispersion effect. 
}
\label{fig:ratio_red_cen_sat_2h}
\end{figure*}


\subsection{Using redshift-space halo power spectra from simulations} \label{sec:sim_model}
In this subsection, we consider the case where we have a perfect model to describe the power spectrum of halos in redshift space, $\widetilde{P}_{XX,h}$ and $\widetilde{P}_{XY,h}$. Specifically, we use measurements from $N$-body simulations for the redshift-space power spectra of halos that have the same bias at large scales and the same number density as the target galaxies.
In section \ref{sec:df_model}, we relax this assumption and use nonlinear perturbation theory to model the halo power spectrum itself.

Central galaxies reside in the centers of their halos and thus, the halo spectrum $P_{cc,h}^S$ in equation \ref{eq:p_cc} can be directly measured from simulations. We compute the halo spectra in equations \ref{eq:p_cas} - \ref{eq:p_sasb} by replacing the position of each satellite galaxy with the center of the halo and assigning the halo velocity to the replaced satellites. This methodology will remove any satellite FoG effects, leaving only RSD due to the halo velocity field.

The measurements from $N$-body simulations for the ratios $P_{c_As}^S/P_{c_As,h}^S$,  $P_{s_As_A}^S/P_{s_As_A,h}^S$, and $P_{s_As_B}^S/P_{s_As_B,h}^S$ are presented as functions of $k$ and $\mu$ in the top panels of figure \ref{fig:ratio_red_cen_sat_2h}, from the left to right, respectively.
If the models in equations \ref{eq:p_cas} - \ref{eq:p_sasb} are correct, the plotted results should be equivalent to the damping factors $G(k\mu; \sigma_{v,s})$, $G(k\mu; \sigma_{v,s_A})^2$, and $G(k\mu; \sigma_{v,s_A})G(k\mu; \sigma_{v,s_B})$, respectively.
For the Gaussian and Lorentzian damping models, we adopt the value of the parameter $\sigma_{v,0}=270 \ {\rm km/s}$ and  $\sigma_{v,0}=210 \ {\rm km/s}$, respectively.
To compute the various velocity dispersions, we use the fact that $\langle  M_{s}^{2/3}\rangle = 2.04\times 10^9(\himsun)^{2/3}$, $\langle  M_{s_A}^{2/3}\rangle = 1.43\times 10^9(\himsun)^{2/3}$, and $\langle  M_{s_B}^{2/3}\rangle = 2.85\times 10^9(\himsun)^{2/3}$, where $M_X$ is the average mass of halos that host satellite galaxy sample $X$. Thus, equation \ref{eq:vel_disp_model} predicts $\sigma_{v,s}=5.67\himpc$, $\sigma_{v,s_A}=4.74\himpc$ and $\sigma_{v,s_B}=6.69\himpc$ for the Gaussian model and $\sigma_{v,s}=4.41\himpc$, $\sigma_{v,s_A}=3.68\himpc$ and $\sigma_{v,s_B}=5.21\himpc$ for the Lorentzian model. The Gaussian and Lorentzian damping models using these velocity dispersions are shown as the dashed and solid lines in the top panels of figure \ref{fig:ratio_red_cen_sat_2h}.

\begin{figure}[bt] 
  \includegraphics[width=\linewidth]{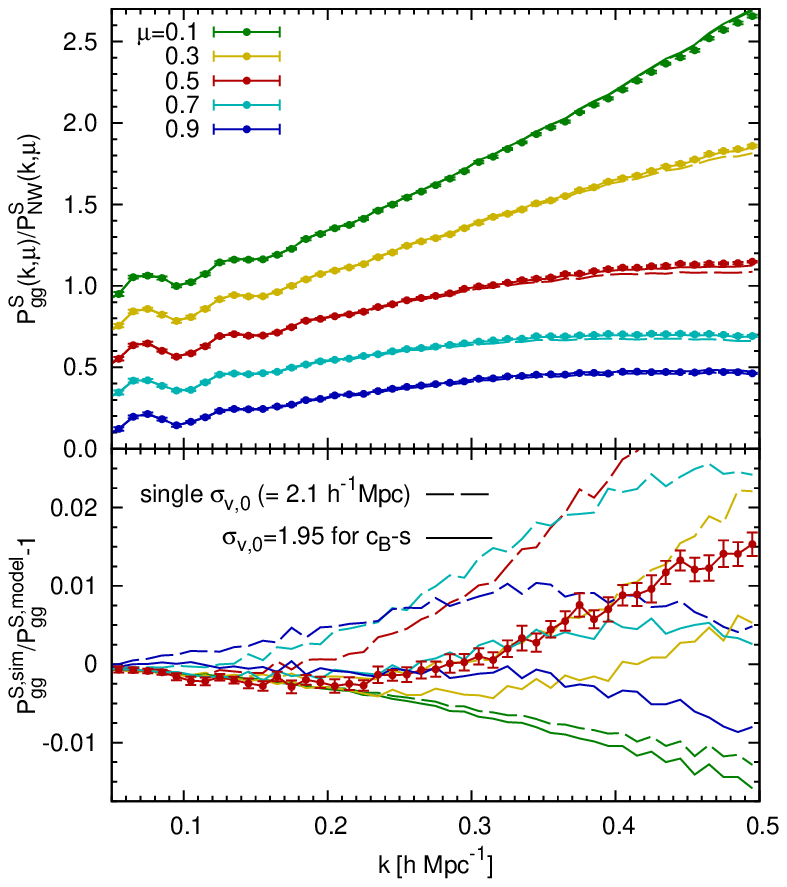}
\caption{{\it Top} : the total galaxy power spectrum in redshift space normalized by the linear Kaiser no-wiggle spectrum in redshift space \cite{Eisenstein:1998}.
The ratio of the redshift-space power spectra for various galaxy samples to the corresponding halo spectra.
The points with error bars are the spectrum directly measured from simulations. 
The dashed lines are our model using halo spectra from simulations with a single parameter of $\sigma_{v,0}=2.1\himpc$, while the solid lines are the case when we adopt $\sigma_{v,0}=1.95\himpc$ for $P_{c_Bs}$. Vertical offsets have been added for clarity.
{\it Bottom} : the ratio of the measured spectrum to our model for the two cases of treating $\sigma_{v,0}$ described above.
For clarity error bars are added only for the $\mu=0.5$ result of the case when the different value of $\sigma_{v,0}$ is adopted for $P_{c_Bs}$.}
\label{fig:ratio_red_sim_gg}
\end{figure}

The Gaussian model fails to explain the small scale RSD of $P_{c_As}^S$ ($k\sim 0.2\hmpci$ for $\mu=0.9$), while the modified Lorentzian model matches the simulation result for all the scales probed here ($k\sim 0.4\hmpci$). 
The results for $P_{s_As_A}^S$ and $P_{s_As_B}^S$ are respectively shown at the middle and right panels of the top set in figure \ref{fig:ratio_red_cen_sat_2h}. The Gaussian model works relatively well for these terms, but the Lorentzian model has a near-perfect agreement with the $N$-body results.

The remaining two terms we wish to model are $P_{c_Bs}^S$ and $P_{s_Bs_B}^S$ which have both 1-halo and 2-halo contributions. 
At the lower set of figure \ref{fig:ratio_red_cen_sat_2h}, we plot $P_{c_Bs}^S/\widetilde{P}_{c_Bs,h}^S$ (left) and $P_{s_Bs_B}^S/[ \tilde{P}_{s_Bs_B}^S-\sigma_{n,s_B}^2 ]$ (right) as functions of $(k,\mu)$. 
For $P_{s_Bs_B}^S$ term, the mass term in equation \ref{eq:vel_disp_model}, $M^{2/3}$, needs to be weighted by the number of satellite pairs in each halo. 
Because the average mass weighted by the number of the pairs is $\langle  M_{s_B}^{2/3}\rangle = 3.78\times 10^9(\himsun)^{2/3}$, 
the $\sigma_{v,0}$ values the same as those for 2-halo terms for Gaussian and Lorentzian models, $G(k\mu;\sigma_{v,s_Bs_B})$, predict $\sigma_{v,s_Bs_B}=7.7\himpc$ and $\sigma_{v,s_Bs_B}=6.0\himpc$, respectively. 
%
%
We find that both the Gaussian and Lorentzian models can explain the measurement of $P_{s_Bs_B}^S$ well. 
On the other hand, we adopt the value of the parameter $\sigma_{v,0}=260 {\rm km/s}$ and  $\sigma_{v,0}=195 {\rm km/s}$, respectively, for the Gaussian and Lorentzian models for $P_{c_Bs}$, which are smaller than those for 2-halo terms by about 7\% and 4\%. 
Since we have a wide halo mass range particularly for the halos which host the central galaxies with satellite(s), $c_B$, it is not surprising that the 1-halo term can have slightly a different value of $\sigma_{v,0}$.

\begin{figure*}[bt]
  \includegraphics[width=.48\linewidth]{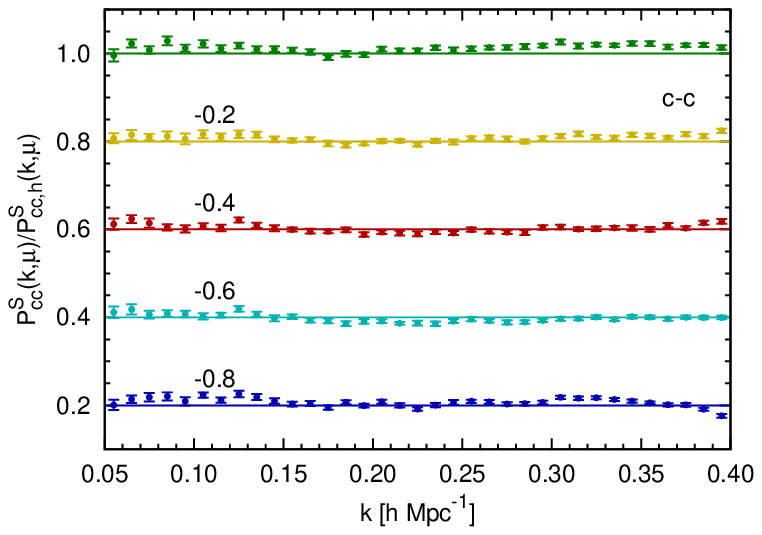}
  \includegraphics[width=.48\linewidth]{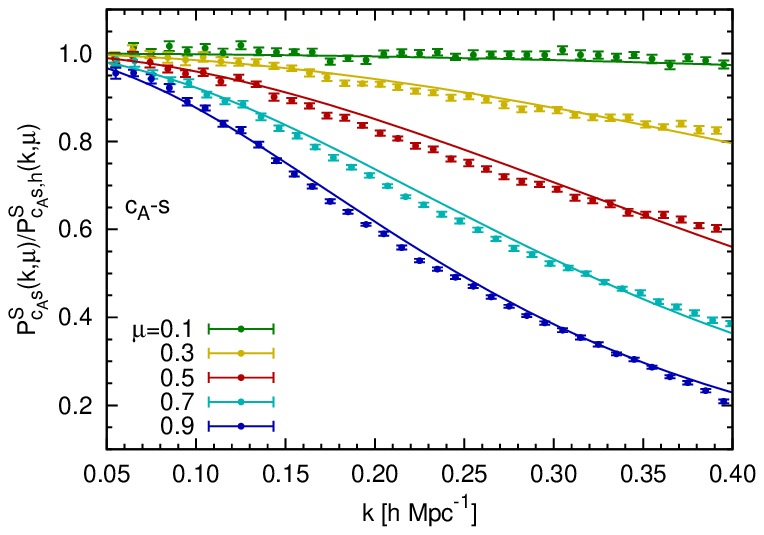}
  \includegraphics[width=.48\linewidth]{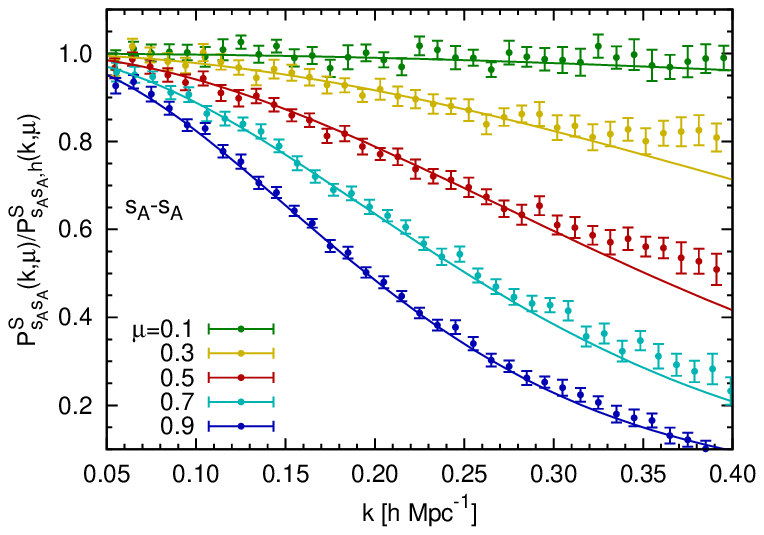}
  \includegraphics[width=.48\linewidth]{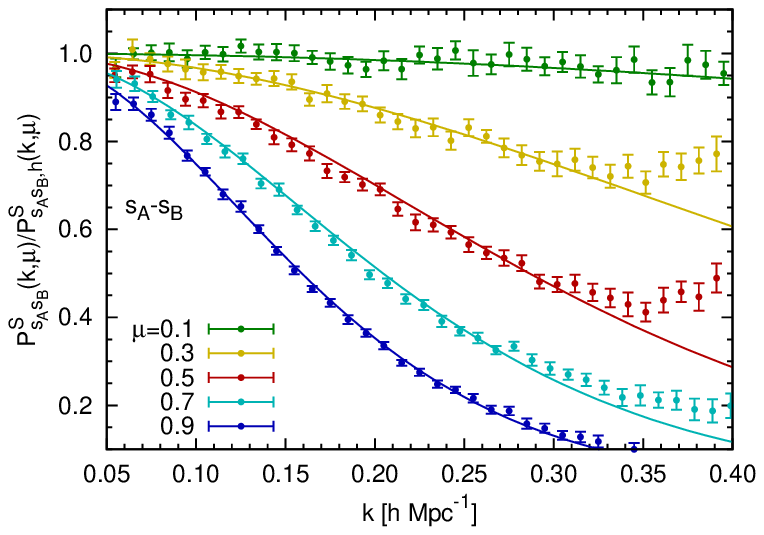}
\caption{The ratio of the 2-halo redshift-space galaxy power spectra to the corresponding halo spectra modeled with the distribution function approach described in section \ref{sec:df_model}. The top left panel shows the result for the auto-spectrum of centrals $(XY)=(cc)$. Centrals have no velocity dispersion, so the expected damping (solid line) is simply unity. For clarity vertical offsets are added for results for different $\mu$ values. The top right panel shows the result for the cross power spectrum of centrals that do not have satellites in the same halo and satellites $(XY)=(c_As)$. The bottom left and right panels show the results for the auto spectrum of satellites that do not have other satellites in the same halos $(XY)=(s_As_A)$ and the cross power between such satellites and satellites that do have satellite(s) in the same halo $(XY)=(s_As_B)$. The solid lines show the predictions for the Lorentzian velocity dispersion damping factor due to the FoG effect (see text for values).}
\label{fig:ratio_red_df_2h}
\end{figure*}

We combine all the above modeling results to see how well our model agrees with the total galaxy power spectrum measured from simulations, $P_{gg}^S$. 
At the top panel of figure \ref{fig:ratio_red_sim_gg}, the points with the error bars show the galaxy power spectrum normalized by the linear power spectrum without baryon acoustic oscillation (BAO) wiggles with the linear Kaiser factor, $P^S_{NW}=(1+f\mu^2/b_{1,g})^2P^R_{NW}$.
The dashed lines the combinations of our modeling results for halo RSD from simulations (equations \ref{eq:p_cc} -- \ref{eq:p_cbs}, \ref{eq:p_sbsb2} --\ref{eq:vel_disp_model}) with a single parameter for the velocity dispersion, $\sigma_{v,0}=2.1\himpc$.
The dashed lines in the bottom panel of figure \ref{fig:ratio_red_sim_gg} show $P_{gg}^{S,sim}/P_{gg}^{S,model}-1$, where $P_{gg}^{S,{\rm sim}}$ and $P_{gg}^{S,{\rm model}}$ are the measured and modeled full galaxy power spectrum, respectively.
We can see that our model is accurate with $2.5\%$ up to $k\sim 0.4\hmpci$.
The solid lines at the top panel of figure \ref{fig:ratio_red_sim_gg} is the same as the dashed lines but we adopt $\sigma_{v,0}=1.95\himpc$ for $P_{c_Bs}$. The points and solid lines with the error bars present the model accuracy, and we achieve the agreement of $\sim 1\%$ at $k\sim 0.4\hmpci$ and $\sim 1.5\%$ at $k\sim 0.5\hmpci$. 


\subsection{Using redshift-space halo power spectrum from perturbation theory} \label{sec:df_model}

In this subsection, we present a model for the galaxy power spectrum in redshift space where we model the halo power spectrum using perturbation theory, rather than measurements from $N$-body simulations as in section \ref{sec:sim_model}. The perturbation theory model used here relies on expressing the redshift-space halo density field in terms of moments of the distribution function (DF), and the approach has been developed and tested in a previous series of papers \cite{Seljak:2011, Okumura:2012, Okumura:2012b, Vlah:2012, Vlah:2013,Blazek:2014}. If we consider halo samples $X$ and $Y$, with linear biases $b_{1, X}$ and $b_{1, Y}$, the redshift-space power spectrum in the DF model is given by

\be
P^S_{XY, h}(k,\mu)=\sum^\infty_{L=0}\sum^\infty_{L'=0}\frac{(-1)^{L'}}{L!L'!} \left(\frac{ik\mu}{\mathcal{H}}\right)^{L+L'} P^{XY,h}_{LL'}(k,\mu),\label{eq:df_power_expansion}
\ee
where $\mathcal{H} = aH$ is the conformal Hubble factor, and $P^{XY,h}_{LL'}$ is the power spectrum of moments $L$ and $L'$ of the radial halo velocity field, weighted by the halo density field. These spectra are defined as 
\be
P^{XY,h}_{LL'}(\vk)(2\pi)^3\delta_D(\vk-\vk')=\left\langle T_\parallel^{X, L}(\vk)T_\parallel^{Y, L'*}(\vk') \right\rangle,
\ee
where $T_\parallel^{X,L}(\vk)$ is the Fourier transform of the halo velocity moments weighted by halo density, $T_\parallel^{X,L}(\vx)=\left[ 1+\delta_X(\vx)\right]v_{\parallel,X}^L$. For example, $P^{XX,h}_{00}$ represents the halo density auto power spectrum of sample $X$, whereas $P^{XX,h}_{01}$ is the cross-correlation of density and radial momentum for halo sample $X$.

The DF approach naturally produces an expansion of $P^S_{XY, h}(k, \mu)$ in even powers of $\mu$, with a finite number of correlators contributing at a given power of $\mu$. For the model presented here, we consider terms up to and including $\mu^4$ order in the expansion of equation \ref{eq:df_power_expansion}. The spectra $P^{XY,h}_{LL'}(k,\mu)$ in equation \ref{eq:df_power_expansion} are defined with respect to the halo field, and a biasing model is needed to relate them to the correlators of the underlying dark matter density field. We use a nonlocal and nonlinear biasing model \cite{Baldauf:2012}, which results in three biasing parameters per halo sample: $b_1$, $b_2^{00}$, and $b_2^{01}$. The spectra $P^{XX, h}_{00}$ and $P^{XX, h}_{01}$ have distinct values for second-order, local bias $b_2$ (see \cite{Vlah:2013} for more details). As discussed in \cite{Vlah:2013}, $b_2^{00}$ and $b_2^{01}$ have a roughly quadratic dependence on the linear bias $b_1$. We fit this dependence to simulations and treat the quadratic biases as a function of $b_1$ only. Thus, the linear bias $b_{1, X}$ is the only free bias parameter for halo sample $X$. Note that when evaluating the cross spectrum $P^S_{XY, h}$ for halo samples $X$ and $Y$, we evaluate the model using the geometric mean of the linear biases, $b_{1, XY} \equiv (b_{1,X}b_{1,Y})^{1/2}$.
As shown in figure \ref{fig:bias}, the power spectra of halos with broad and narrow mass distributions (denoted as "c" and "bin2", respectively) 
have slightly different bias but very similar shapes, confirming that one can replace a broad mass distribution of centrals 
with a narrow mass distribution. This may not be such a good approximation for satellites, but if the satellite fraction is low 
the overall effects are small as well.  

To evaluate the underlying dark matter correlators and nonlinear biasing terms present in equation \ref{eq:df_power_expansion}, we use Eulerian perturbation theory, as described in detail in \cite{Vlah:2013}. Perturbation theory breaks down on small scales, and in order to increase the accuracy of the DF model, we use results calibrated from simulations in three instances. First, we use the Halo Zeldovich model \cite{Seljak:2015} for the dark matter correlators $P_{00}$ and $P_{01}$, where $P_{01}$ can be related to $P_{00}$ through $P_{01} = \mu^2 \mathrm{d}P_{00} / \mathrm{dln}a$ \cite{Vlah:2012}. This model has been shown by \cite{Seljak:2015} to be accurate to 1\% to $k = 1 \ihmpc$. Second, we use simulation measurements for the functional forms of the cross-correlation between dark matter density and velocity divergence, $P_{\delta \theta}$, and the $\mu^4$-dependence of the dark matter momentum density auto-correlation, $P_{11}[\mu^4]$. Specifically, $P_{11}[\mu^4] / P_\mathrm{lin}$ and $P_{\delta \theta}/P_\mathrm{lin}$ are interpolated from simulations as a function of $f^2\sigma_8^2$ and $f\sigma_8^2$, respectively. Finally, there are corrections to the Poisson model for shot noise $\sigma^2_{n,X}$ due to halo exclusion effects and nonlinear clustering \cite{Baldauf:2013, Vlah:2013}. As first shown in \cite{Baldauf:2013}, these two corrections to the halo stochasticity must be considered together, with exclusion leading to a suppression of power in the low-$k$ limit and nonlinear clustering providing an enhancement. The deviations from Poisson stochasticity are typically a few percent in the low-$k$ limit and must vanish in the high-$k$ limit, leading to a complicated scale-dependence. In previous work, \cite{Vlah:2013} modeled the halo stochasticity (denoted as $\Lambda$ in \cite{Vlah:2013}) with an ad-hoc fitting formula, which worked well over the desired range of scales. Rather than use this fitting formula, we measure the halo stochasticity directly from simulations over a range of redshifts and halo mass bins, and then interpolate the results as a function of linear bias. The results
interpolated directly from simulations described here are designed to be accurate to a maximum wavenumber of $k = 0.4 \ihmpc$. Improved theoretical modeling, independent of simulation measurements, is actively being developed.

We can compute the redshift-space halo spectra $P^S_{XY, h}$ in the four 2-halo galaxy power spectra of equations \ref{eq:p_cc} -- \ref{eq:p_sasb} using the DF approach described in this section. Note that $P_{c_A s, h}^S$ and $P_{s_As_B, h}^S$ are cross power spectra of two halo samples, and we model the spectra using the geometric mean of the linear biases of the individual samples. The ratio of the simulation measurements for these four 2-halo galaxy spectra to the corresponding halo spectra modeled with the DF approach in redshift space is shown in figure \ref{fig:ratio_red_df_2h}. For $P_{c_As}^S$, $P_{s_A s_A}^S$, and $P_{s_A s_B}^S$, the solid lines show the suppression factor due to the FoG effect using the Lorentzian model. No FoG suppression is expected for $P_{cc}$ so the solid lines are simply unity. To compute the halo spectra, the linear biases are taken from table \ref{tab:gal_halo}, and we vary the FoG velocity
dispersions to obtain the best fit possible. The values we obtain are $\sigma_{v,s} = 4.1\himpc$, $\sigma_{v,s_A}=3.5\himpc$, and $\sigma_{v,s_B}=5\himpc$. These values are in rough agreement with those predicted using equation \ref{eq:vel_disp_model}, with discrepancies likely due to deficiencies in the RSD model at high $k$ and high $\mu$, which are partially compensated by the FoG damping. 

\begin{figure}[tb]
  \includegraphics[width=\linewidth]{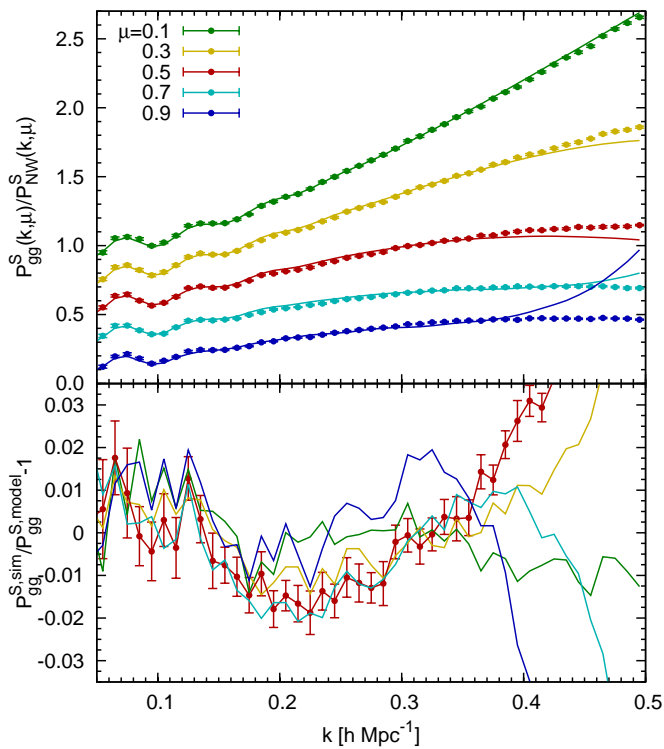}
\caption{{\it Top} : the total redshift-space galaxy power spectrum $P^S_{gg}(k,\mu)$ as measured from simulations (points with errors). The solid lines show the prediction for $P^S_{gg}$ using the distribution function model for the halo spectra of the various 2-halo terms contributing to the total power. The power spectra have been normalized by the linear Kaiser redshift space power spectrum, using the the no-wiggle linear power spectrum \cite{Eisenstein:1998}. Vertical offsets have been added for clarity. {\it Bottom} : the ratio of the measured spectrum to our DF model. For clarity the error bars are added only for the $\mu=0.5$ result. }\label{fig:red_df_total}
\end{figure}

We can also combine both 2-halo and 1-halo terms in order to examine the accuracy of the DF model in describing the total galaxy redshift-space power spectrum, $P^S_{gg}(k, \mu)$. Similar to section \ref{sec:sim_model}, we use equations \ref{eq:p_cbs} and \ref{eq:p_sBsB} to model $P^S_{c_Bs}$ and $P^S_{s_B s_B}$, and we use the DF model to describe the halo spectra that enter into these equations. The FoG velocity dispersions used for these terms are the same as described previously: $\sigma_{v,s} = 4.1\himpc$ and $\sigma_{v,s_B}=5\himpc$. The same values are used for both the 1-halo and 2-halo contributions. The amplitudes of the 1-halo terms for $P^S_{c_Bs}$ and $P^S_{s_B s_B}$ are treated as constant and equal to $\Sigma_{c_B}^2$ and $\Sigma^2_{s_B}$, respectively. We show the simulation measurements for $P_{gg}^S(k, \mu)$ (points with errors) as well as the DF model plus FoG damping prediction (solid lines) at the top panel of figure \ref{fig:red_df_total}. The power spectra in this figure have been normalized by the linear Kaiser redshift space power spectrum, using the the no-wiggle linear power spectrum \cite{Eisenstein:1998}. 
The bottom panel of figure \ref{fig:red_df_total} shows the accuracy of our DF model compared to the measurement, $P_{gg}^{S,sim}/P_{gg}^{S,model}-1$.
Using the DF model described here for halo spectra, we can successfully describe the total galaxy spectrum in redshift space to small scales, roughly $k \sim 0.4 \ihmpc$ within $3\%$.
The modeled spectrum breaks down significantly at $k>0.4\himpc$, because the underlying PT breaks down at these scales and the simulation-calibrated results have only been fit to $k\leq 0.4\himpc$ in order to improve the accuracy only to this wavenumber. 

The results shown in figure \ref{fig:red_df_total} use three free parameters, given by the three FoG velocity dispersions $\sigma_{v,s}$, $\sigma_{v,s_A}$, and $\sigma_{v,s_B}$; all other parameters (linear biases, sample fractions, etc) are set to their fiducial values listed in table \ref{tab:gal_halo}. In comparison to the results of \ref{sec:sim_model}, where simulation measurements are used for halo spectra, the DF model achieves a slightly worse precision (1\% vs. 3\% at $k \sim 0.4 \ihmpc$) and uses one additional free parameter to model the FoG velocity dispersions.

\section{Conclusions}\label{sec:conclusion}

In this paper we have investigated the redshift-space power spectrum of galaxies using $N$-body simulations and developed a model based on
perturbation theory (PT) of dark matter halos. 
In previous work \cite{Vlah:2013} we have established the requirements and reach of PT models on halos, and in this paper our focus 
is on effects induced by satellites that are not at the halo centers, inducing effects that 
go beyond PT halo modeling. We have argued that the simplest approach to describe these effects is within the context of a halo model, 
where the radial distribution of satellites inside the halo induces both 2-halo and 1-halo effects. 
In real space these effects add additional small scale clustering term, the so called 1-halo term, 
which appears at low $k$ as a white noise like term. We have investigated the departures of this term from the white noise
and found that they can be well approximated as a $-k^2R^2$  relative
correction to the 1-halo power spectrum. The corresponding 2-halo term effects are also 
very small and scale as $-k^2R^2P_L(k)$. Together the two effects can be modeled as $-k^2R^2P(k)$ correction, 
but are in any case very small. 

In redshift space the satellites are 
spread out in the radial direction 
by their virial velocities inside the halos, an effect called Fingers-of-God (FoG). 
These FoG effects also induce both 1-halo and 2-halo correlations. These effects are large and over the range of 
scales of interest they cannot be modeled simply by a $k^2\sigma_v^2$ correction. Instead, we explore several FoG resummations
proposed in the literature, finding that Lorentzian is a good fit over the range of interest.  
To provide a more 
physical interpretation we further decompose these galaxy subsamples into subsamples that can be 
approximately described as having a narrow halo mass distribution. 
The advantage of this decomposition is that there is just a single FoG term that is needed to describe 
both the 1-halo and 2-halo FoG, and that this term is related to the typical virial velocity corresponding to the halos of a given mass. 
We divide
centrals into those with and without satellite inside the same halos, and satellites into those with and without another satellite(s) inside the same halos. The decomposed terms of the observed power spectrum can be uniquely related to 1-halo and 2-halo terms in a halo model and we assign each subsample 
their own FoG term. But doing this we successfully model 
the contributions up to $k \sim 0.4 \hmpci$, and we can relate the FoG parameters to the underlying physical 
properties of the halos like their halo mass. 

Ultimately our goal is to model the observed galaxy power spectrum from surveys such as BOSS \cite{Schlegel:2009, Eisenstein:2011}. Our 
modeling contains many more parameters than the current state-of-the art models of RSD power spectrum, 
which typically combine PT halo models 
with a single white noise amplitude and a single FoG term to account for satellite effects 
\cite{Beutler:2014}. These parameters combine all the different terms discussed here into a single one, and as 
we argued there should be several FoG terms that act differently on different scales. As a 
consequence these models typically fail for $k>0.2\hmpci$, while our models extend the reach up to 
$k \sim 0.4\hmpci$. Moreover, our FoG parameters can be directly connected to the underlying 
halo mass of halos in which satellites live, and thus the bias of the same halos, 
and our 1-halo amplitude can be connected to the satellite fraction, so many of the parameters we 
introduced may have strong priors and do not need to be fit from the data. There is 
thus hope that such improved modeling of small scales will translate into better cosmological 
constraints, as argued recently for
configuration space analysis \cite{Reid:2014}, and we hope to address this in the future. 

\section*{Acknowledgment} 
T.O. thanks Masahiro Takada for discussion. We also thank the referee for useful comments which improve the presentation of this paper.
T.O. is supported by Grant-in-Aid for Young Scientists (Start-up) from the Japan Society for the Promotion of Science (JSPS) (No. 26887012).
N.H. is supported by the National Science Foundation Graduate Research Fellowship under Grant No. DGE-1106400 and the Berkeley Fellowship for Graduate Study. 
U.S. is supported in part by the NASA ATP grant NNX12AG71G.
Z.V. is supported in part by the U.S. Department of Energy contract to SLAC no. DE-AC02-76SF00515.

\bibliography{ms.bbl}
 \end{document}